\documentclass[a4]{article}
\usepackage[T1]{fontenc}
\usepackage[latin1]{inputenc}
\usepackage{amsmath, amssymb, amsthm}
\usepackage[english]{babel}
\usepackage{indentfirst}
\usepackage{booktabs}
\usepackage{array}
\usepackage{caption,appendix}
\usepackage{geometry}
\usepackage{float}
\usepackage{cancel}
\usepackage{bbold}
\usepackage{authblk}
\numberwithin{equation}{section}
\begin{document}
\author[1,2,3]{Salvatore Capozziello \thanks{capozziello@na.infn.it}}
\author[4]{Maurizio Capriolo \thanks{mcapriolo@unisa.it} }
\affil[1]{\emph{Dipartimento di Fisica "E. Pancini", Universit\`a di Napoli {}``Federico II'', Compl. Univ. di
		   Monte S. Angelo, Edificio G, Via Cinthia, I-80126, Napoli, Italy, }}
		  \affil[2]{\emph{INFN Sezione  di Napoli, Compl. Univ. di
		   Monte S. Angelo, Edificio G, Via Cinthia, I-80126,  Napoli, Italy,}}
 \affil[3]{\emph{ Lab. Theor. Cosmology, Tomsk State University of Control Systems and Radioelectronics (TUSUR), 634050 Tomsk, Russia.}}
\affil[4]{\emph{Dipartimento di Matematica Universit\`a di Salerno, via Giovanni Paolo II, 132, Fisciano, SA I-84084, Italy.} }
\date{\today}
\title{\textbf{Gravitational waves in non-local gravity}}
\maketitle
\begin{abstract} 
We derive  gravitational waves in a theory with non-local  curvature corrections to the  Hilbert-Einstein Lagrangian.  In addition to the standard  two massless tensor modes,  with plus and cross polarizations,   helicity 2 and angular frequency $\omega_{1}$,   we obtain a further  scalar   massive mode with helicity 0 and angular frequency $\omega_{2}$,  whose polarization is  transverse. It is  a breathing mode,  which, at  the lowest order of an effective  parameter $\gamma$,    presents a speed difference  between nearly null and null plane waves.  Finally,  the quasi-Lorentz  $E(2)$-invariant class for the non-local  gravity is type $N_{3}$,  according to the Petrov classification.    This means that  the presence (or absence) of gravitational wave modes is observer-independent.
\end{abstract}



\section{Introduction}
Theories of physics describing elementary interactions are local, that is, fields are  evaluated at the same point,  and are governed by point-like Lagrangians from which one derives equations of motion.  However,  already at  classical level,  it is possible to observe  non-locality in Electrodynamics  of continuous media,  when  spatial or temporal dispersions, due to the non-local constitutive relation between the fields $\left(\mathbf{D}, \mathbf{H}\right)$ and $\left(\mathbf{E},\mathbf{B}\right)$, occur \cite{LLECM, JED, MOP, CM} as
\begin{equation}
D_{i}\left(x\right)=\int\,d^{4}x^{\prime}\epsilon_{ij}\left(x^{\prime}\right)E^{j}\left(x-x^{\prime}\right).
\end{equation}
It is a sort of memory-dependent phenomenon  taking into account both the past history of the fields and their values taken in other points of the medium.  Also at  quantum level, some effective  actions show  non-local terms and therefore, the  associated field equations are integro-differential ones.  Recently, non-locality has been considered in cosmology taking into account  non-local models to explain early and late-time cosmic acceleration as well as  structure formation, without introducing  dark energy and  dark matter.  Specifically, non-locality can play interesting roles to address  problems like   cosmological constant,   Big Bang and black hole singularities,  and, in general,  coincidence and fine-tuning problems, which affect the  $\Lambda$CDM model  \cite{NO, DWCI, MAMA, BDFM, SDSP, NCA}.  

Some  non-local field theories are of infinite order because they have an infinite number of derivatives. This feature is  due to the presence of operators like $f\left(\Box\right)$ which  can be expanded in series, assuming that $f$ is an analytic function,  as 
\begin{equation}
f(\Box)=\sum_{n=0}^{\infty}a_{n}\Box^{n}\ .
\end{equation}
Here $\Box$ is the d'Alembert operator. This procedure is aimed  to make the theory ghost free \cite{BLM, BKLM, BKLMM,  BLY}.  

In particular, we know that General Relativity describes gravity as a local interaction while  Quantum Mechanics shows non-local aspects. Several approaches have been proposed to achieve a self-consistent Quantum Gravity as discussed, for example, in  \cite{CALC, Petrov}.  A possibility towards Quantum Gravity is   considering  non-local corrections to the Hilbert-Einstein action \cite{Modesto1,Modesto2}. It is a natural way to cure  ultraviolet and infrared behaviors of General Relativity. Introducing non-local terms can work also in alternative theories like   teleparallel gravity   \cite{BCFETALL}.  
 
 In all these approaches, it is important to  study the linearized versions of the theories   and to  derive  gravitational waves (GWs). In fact, gravitational radiation allows to detect possible effects of non-local gravity \cite{CCN} as well as to classify the degrees of freedom of a given theory \cite{GW17_1, GW17_2, GW18}.

In this paper, we want to investigate the effect of non-locality in a  theory of gravity where the Ricci scalar $R$ of General Relativity is corrected by  $R\Box^{-1}R$,  which is the first interesting non-local curvature term.  Specifically, we want to investigate how   polarization, helicity, and mass of GWs are affected by this kind of  terms  and how the  $E(2)$ Petrov classification changes.

 Sec.  \ref{a}  is devoted to a procedure for the  localization  of the  non-local gravitational action by the method of Lagrangian multipliers. Starting from this localized action,   it is possible to derive  the field equations.   Subsequently, in Sec. \ref{b}, we linearize the field equations and  solve them in   harmonic gauge and in a further  gauge in view of  eliminating   ghost modes. Finally,   we  get the GWs.  In Sec.\ref{c} and \ref{d},  polarizations are analyzed by using both geodesic deviation  and the Newman-Penrose formalism. The method consists in   expanding the nearly null massive plane waves in term of the exactly null plane waves. The expansion is achieved in terms of a parameter $\gamma_{k}$.  In  Sec.  \ref{e}, results are summarized and  possible future developments are discussed.

\section{Localization of non-local gravity action via Lagrange multipliers approach}\label{a}

Let us study gravitational interaction  governed by the following non-local action
\begin{equation}\label{1}
S[g]=\frac{1}{2\kappa^2}\int d^4x \sqrt{-g} \left(R+a_{1}R\Box^{-1}R\right)+\int d^4x \sqrt{-g}\mathcal{L}_{m}[g]\ ,
\end{equation}
where $k^{2}=8\pi G/c^{4}$.  Its field equations are non-linear integro-differential equations  due to the non-local term.  We introduce the auxiliary field $\phi(x)$ defined as 
\begin{equation}\label{2}
\phi(x)=\Box^{-1}R\ ,
\end{equation}
and then the Ricci scalar is
\begin{equation}\label{3}
R=\Box\phi(x)\,.
\end{equation}
According to this definition, a Lagrange multiplier can be considered    so that the  gravitational action becomes \cite{NO}
\begin{equation}\label{4}
S_{g}[g,\phi,\lambda]=\frac{1}{2\kappa^2}\int d^4x \sqrt{-g} \left[R\left(1+a_{1}\phi\right)+\lambda\left(\Box\phi-R\right)\right]\, ,
\end{equation}
where $\lambda(x)$ is a further scalar field.
Using integration by parts  and imposing that fields and their derivatives  vanish onto the boundary of integration domain, we obtain
\begin{equation}\label{5}
S_{g}[g,\phi,\lambda]=\frac{1}{2\kappa^2}\int d^4x \sqrt{-g} \left[R\left(1+a_{1}\phi-\lambda\right)-\nabla^{\nu}\lambda\nabla_{\nu}\phi\right]\ .
\end{equation}
Varying with respect to $\phi$,  we get
\begin{equation}\label{6}
\frac{1}{\sqrt{-g}}\frac{\delta S}{\delta\phi}=\frac{1}{2\kappa^{2}}\left[a_{1}R+\Box\lambda\right]=0\Rightarrow \Box\lambda=-a_{1}R\ ,
\end{equation}
while varying with respect to $\lambda$,  the functional derivative takes the  form 
\begin{equation}\label{7}
\frac{1}{\sqrt{-g}}\frac{\delta S}{\delta\lambda}=\frac{1}{2\kappa^{2}}\left[-R+\Box\phi\right]=0\Rightarrow\Box\phi=R\ .
\end{equation}
Finally,  the variation with respect to the metric $g^{\mu\nu}$ gives 
\begin{equation}\label{8}
\frac{2\kappa^{2}}{\sqrt{-g}}\frac{\delta S_{g}}{\delta g^{\mu\nu}}=\left(G_{\mu\nu}+g_{\mu\nu}\Box-\nabla_{\mu}\nabla_{\nu}\right)\left(1+a_{1}\phi-\lambda\right)-\nabla_{(\mu}\phi\nabla_{\nu)}\lambda+\frac{1}{2}g_{\mu\nu}\nabla^{\sigma}\phi\nabla_{\sigma}\lambda\ ,
\end{equation}
and 
\begin{equation}\label{9}
\frac{\delta\left(\sqrt{-g}\mathcal{L}_{m}\right)}{\delta g^{\mu\nu}}=-\frac{\sqrt{-g}}{2}T_{\mu\nu}\ .
\end{equation}
The final field equations are 
\begin{equation}\label{eqfieldginmatter}
\boxed{
\left(G_{\mu\nu}+g_{\mu\nu}\Box-\nabla_{\mu}\nabla_{\nu}\right)\left(1+a_{1}\phi-\lambda\right)-\nabla_{(\mu}\phi\nabla_{\nu)}\lambda+\frac{1}{2}g_{\mu\nu}\nabla^{\sigma}\phi\nabla_{\sigma}\lambda=\kappa^{2}T_{\mu\nu}
}\ ,
\end{equation}
\begin{equation}\label{eqfieldphi}
\boxed{
\Box\phi=R
}\ ,
\end{equation}
\begin{equation}\label{eqfieldlambda}
\boxed{
\Box\lambda=-a_{1}R
}
\end{equation}\label{10}
where $G_{\mu\nu}$ is the Einstein tensor
\begin{equation}\label{10.1}
G_{\mu\nu}=R_{\mu\nu}-\frac{1}{2}\eta_{\mu\nu}R\ .
\end{equation}
According to Eqs. \eqref{eqfieldphi} and \eqref{eqfieldlambda}, the trace of Eq.  \eqref{eqfieldginmatter} is
\begin{equation}\label{eqtrace}
\boxed{
\left[1+a_{1}\left(\phi-6\right)-\lambda\right]R-\nabla^{\rho}\phi\nabla_{\rho}\lambda=-\kappa^{2}T
}\,,
\end{equation}
and now we can start our considerations on the weak field behavior of this theory.

\section{The weak filed limit  and gravitational waves}\label{b}

In order to analyze gravitational radiation, let us perturb the metric tensor $g_{\mu\nu}$ around the flat metric $\eta_{\mu\nu}$ and the two scalar fields $\phi$ and $\lambda$ around their  values in  Minkowskian spacetime $\phi_{0}$ and $\lambda_{0}$. It is  
\begin{align}\label{11}
g_{\mu\nu}\sim\eta_{\mu\nu}+h_{\mu\nu}\ ,\\ 
\phi\sim\phi_{0}+\delta\phi\ ,\\ 
\lambda\sim\lambda_{0}+\delta\lambda\ .
\end{align}
At first order in $h_{\mu\nu}$, the Ricci  tensor $R_{\mu\nu}$ and the Ricci scalar $R$ become 
\begin{equation}\label{12}
R_{\mu\nu}^{(1)}=\frac{1}{2}\left(\partial_{\sigma}\partial_{\mu}h^{\sigma}_{\nu}+\partial_{\sigma}\partial_{\nu}h^{\sigma}_{\mu}-\partial_{\mu}\partial_{\nu}h-\Box h_{\mu\nu}\right)\ ,
\end{equation}
\begin{equation}\label{13}
R^{(1)}=\partial_{\mu}\partial_{\nu}h^{\mu\nu}-\Box h_{\mu\nu}\ ,
\end{equation}
where $h$ is the trace of perturbation $h_{\mu\nu}$.
In vacuum and under Lorentz gauge, Eqs.  \eqref{12}, \eqref{13} and the Einstein tensor $G_{\mu\nu}$ become
\begin{equation}\label{14}
R_{\mu\nu}^{(1)}=-\frac{1}{2}\Box h_{\mu\nu}\ ,
\end{equation}
\begin{equation}\label{15}
R^{(1)}=-\frac{1}{2}\Box h\ ,
\end{equation}
\begin{equation}\label{16}
G_{\mu\nu}^{(1)}=-\frac{1}{2}\Box h_{\mu\nu}+\frac{1}{4}\eta_{\mu\nu}\Box h\ .
\end{equation}
Later,  according to Eqs. \eqref{eqfieldginmatter},  \eqref{eqfieldphi} and \eqref{eqfieldlambda},  the linearized field equations in vacuum are  
\begin{equation}\label{eqling}
\left(1+a_{1}\phi_{0}-\lambda_{0}\right)\left(2\Box h_{\mu\nu}-\eta_{\mu\nu}\Box h\right)+4\partial_{\mu}\partial_{\nu}\left(a_{1}\delta\phi-\delta\lambda\right)+4a_{1}\eta_{\mu\nu}\Box h=0\ ,
\end{equation}
\begin{equation}\label{eqlinphi}
\Box\left(2\delta\phi+h\right)=0\ ,
\end{equation}
\begin{equation}\label{eqlinlambda}
\Box\left(2\delta\lambda-a_{1}h\right)=0\ ,
\end{equation}
and, from Eqs. \eqref{eqlinphi} and \eqref{eqlinlambda},  the linearized trace equation is
\begin{equation}\label{Eqlineartrace}
\left[1+a_{1}\left(\phi_{0}-6\right)-\lambda_{0}\right]\Box h=0\ .
\end{equation}
The trace equation \eqref{Eqlineartrace} admits solutions if
\begin{equation}\label{caseA}
1+a_{1}\phi_{0}-\lambda_{0}\neq 6a_{1}\quad\rightarrow\quad\Box h=0\ ,
\end{equation}
or
\begin{equation}\label{caseB}
1+a_{1}\phi_{0}-\lambda_{0}=6a_{1}\quad\rightarrow\quad\Box h\neq 0\quad\text{and}\quad\Box h=0\ .
\end{equation}
In the case \eqref{caseA},  Eqs. \eqref{eqling}, \eqref{eqlinphi} and \eqref{eqlinlambda} become
\begin{equation}\label{eqgcaseAC}
\left(1+a_{1}\phi_{0}-\lambda_{0}\right)\Box h_{\mu\nu}+2\partial_{\mu}\partial_{\nu}\left(a_{1}\delta\phi-\delta\lambda\right)=0\ ,
\end{equation}
\begin{equation}\label{eqphicaseAC}
\Box\delta\phi=0\ ,
\end{equation}
\begin{equation}\label{eqlambdacaseAC}
\Box\delta\lambda=0\ .
\end{equation}
In $k$-space, considering the Fourier transform, Eq. \eqref{eqgcaseAC} takes the following form
\begin{equation}\label{eq1}
\left(1+a_{1}\phi_{0}-\lambda_{0}\right)k^{2}\tilde{h}_{\mu\nu}\left(k\right)+2k_{\mu}k_{\nu}\left(a_{1}\tilde{\delta\phi}\left(k\right)-\tilde{\delta\lambda}\left(k\right)\right)=0\,.
\end{equation}
It implies
\begin{equation}\label{eq2}
\Box h=0\Rightarrow k^{2}\tilde{h}\left(k\right)=0\Rightarrow k^2=0\ ,
\end{equation}
and, putting  \eqref{eq2} into \eqref{eq1},  we get a  solution if
\begin{equation}\label{eq2a}
a_{1}\tilde{\delta\phi}\left(k\right)=\tilde{\delta\lambda}\left(k\right)\ ,
\end{equation}
which,  in $x$-space,  becomes 
\begin{equation}\label{eq2aa}
a_{1}\delta\phi\left(x\right)=\delta\lambda\left(x\right)\ .
\end{equation}
Inserting \eqref{eq2aa} into \eqref{eqgcaseAC},  for $1+a_{1}\phi_{0}-\lambda_{0}\neq 0$,  we obtain 
\begin{equation}
\Box h_{\mu\nu}=0\ ,
\end{equation}
which,  together with Eqs. \eqref{eqphicaseAC} and \eqref{eqlambdacaseAC},  yield massless, 2-helicity transverse waves solutions for $k_{1}^{2}=0$,  namely the standard gravitational waves of General Relativity.
In the case \eqref{caseB},  excluding  $\Box h=0$ because the previous massless case returns,  Eqs. \eqref{eqling}, \eqref{eqlinphi} and \eqref{eqlinlambda} become a coupled partial differential equations system,  i.e.
\begin{equation}\label{eqgcaseB}
6a_{1}\Box h_{\mu\nu}-a_{1}\eta_{\mu\nu}\Box h+2\partial_{\mu}\partial_{\nu}\left(a_{1}\delta\phi-\delta\lambda\right)=0\ ,
\end{equation}
\begin{equation}\label{eqphicaseB}
\Box\delta\phi=-\frac{1}{2}\Box h\ ,
\end{equation}
\begin{equation}\label{eqlambdacaseB}
\Box\delta\lambda=\frac{a_{1}}{2}\Box h\ ,
\end{equation}
with the additional condition
\begin{equation}\label{nablahneq0}
\Box h\neq 0\ .
\end{equation}
From Eq. \eqref{nablahneq0}, in the momentum space,  we have
\begin{equation}
k_{2}^{2}\tilde{h}\left(k\right)\neq 0\Rightarrow k_{2}^{2}\neq 0\quad \text{and} \quad \tilde{h}\left(k\right)\neq 0\ ,
\end{equation}
where $\left(k_{2}\right)^{\mu}=(\omega_{2},\mathbf{k})$ is the wave four-vector and $k_{2}^{2}=M^{2}$.
We assume that Eq. \eqref{nablahneq0} is 
\begin{equation}\label{eq3}
\Box h\left(x\right)=g\left(x\right)\ ,
\end{equation}
choosing $g(x)$ with any $k^{2}\neq0$ such as 
\begin{align}\label{possiblefunctiong}
g(x)=&\frac{1}{(2\pi)^{3/2}}\int d^{3}\mathbf{k}\;\tilde{q}\left(\mathbf{k}\right)e^{i\varphi}e^{i\omega_{2}t}e^{-i\mathbf{k}\cdot \mathbf{x}}+c.c.\\=&\frac{1}{(2\pi)^{3/2}}\int d^{3}\mathbf{k}\;2\tilde{q}\left(\mathbf{k}\right)\cos\left(\omega_{2}t-\mathbf{k}\cdot\mathbf{x}+\varphi\right)\ ,
\end{align}
where $\tilde{q}\left(\mathbf{k}\right)\in L^{2}\left(\mathbf{R}^{3}\right)$ is a square integrable function in $\mathbf{R}^{3}$ and $\varphi$ is the phase.  By performing the Fourier transform with respect to the spatial coordinates,  the non-homogeneous wave equation \eqref{eq3} becomes, in $\mathbf{k}$-space,  
\begin{equation}
\left(\partial_{0}^{2}+|\mathbf{k}|^{2}\right)\tilde{h}\left(t,\mathbf{k}\right)=\tilde{q}\left(\mathbf{k}\right)e^{i\varphi}e^{i\omega_{2}t}\ ,
\end{equation}
which has a particular solution
\begin{equation}\label{eq4}
\tilde{h}\left(t,\mathbf{k}\right)=-\frac{\tilde{q}\left(\mathbf{k}\right)e^{i\varphi}}{k_{2}^{2}}e^{i\omega_{2}t}=\tilde{A}\left(\mathbf{k}\right)e^{i\omega_{2}t}\ ,
\end{equation}
where $\omega_{2}=\sqrt{M^{2}+|\mathbf{k}|^{2}}$. The value $M^{2}$ corresponds to  the squared mass of a new  effective   scalar field $\Psi$ which can be defined as the combination
\begin{equation}
\Psi\left(x\right)=\frac{\lambda\left(x\right)}{a_{1}}-\phi\left(x\right),
\end{equation}
whose linear perturbation $\delta \Psi$ satisfies the Klein-Gordon equation
\begin{equation}
\left(\Box +M^{2}\right)\delta \Psi=0,
\end{equation}
under the assumptions that $g(x)$ behaves as  a wave packet with $k^{2}$ exactly equal to  $M^2$.  
It is worth noticing that the  linear perturbation  of $\Psi$ in $\mathbf{k}$-space, $\delta\tilde{\Psi}(\mathbf{k})$, corresponds to $\tilde{h}(\mathbf{k})=\tilde{A}(\mathbf{k})$ see Eq. \eqref{eq4}. It is the trace of metric perturbation in $\mathbf{k}$-space.   From Eqs.  \eqref{eqphicaseB}, \eqref{eqlambdacaseB} and \eqref{eq3}, it is
\begin{equation}
\Box \Psi\left(x\right)=\Box \left(\frac{\lambda\left(x\right)}{a_{1}}-\phi\left(x\right)\right)=\Box h,
\end{equation}
which, in $\mathbf{k}$-space with non-null $k^2$, becomes 
\begin{equation}
\delta\tilde{\Psi}\left(\mathbf{k}\right)=\frac{\delta\tilde{\lambda}\left(\mathbf{k}\right)}{a_{1}}-\delta\tilde{\phi}\left(\mathbf{k}\right)=\tilde{A}\left(\mathbf{k}\right)\ .
\end{equation}
With these considerations in mind, a solution of Eq. \eqref{eq3} is
\begin{equation}\label{eq5}
h(x)=\frac{1}{(2\pi)^{3/2}}\int d^{3}\mathbf{k}\;\tilde{A}\left(\mathbf{k}\right)e^{ik_{2}\cdot x}+ c.c.\ ,
\end{equation}
and, according to the two wave Eqs. \eqref{eqphicaseB}, \eqref{eqlambdacaseB} in $\mathbf{k}$-space, it is
\begin{equation} \label{e6}
\left(\partial_{0}^{2}+|\mathbf{k}|^{2}\right)\left[2\,\tilde{\delta\phi}\left(t,\mathbf{k}\right)+\tilde{h}\left(t,\mathbf{k}\right)\right]=0\ ,
\end{equation}
\begin{equation}\label{eq7}
\left(\partial_{0}^{2}+|\mathbf{k}|^{2}\right)\left[2\,\tilde{\delta\lambda}\left(t,\mathbf{k}\right)-a_{1}\tilde{h}\left(t,\mathbf{k}\right)\right]=0\ .
\end{equation}
From Eq. \eqref{eq4},  we obtain two particular solutions for fixed $\mathbf{k}$ 
\begin{equation}\label{eq8}
\tilde{\delta\phi}\left(t,\mathbf{k}\right)=-\frac{1}{2}\tilde{A}\left(\mathbf{k}\right)e^{i\omega_{2}t}\ ,
\end{equation}
\begin{equation}\label{eq9}
\tilde{\delta\lambda}\left(t,\mathbf{k}\right)=\frac{a_{1}}{2}\tilde{A}\left(\mathbf{k}\right)e^{i\omega_{2}t}\ ,
\end{equation}
that allows us to derive the waves $k_{1}^{2}=0$ and $k_{2}^{2}=M^{2}$ for the two scalar linear perturbations $\delta\phi$ and $\delta\lambda$, i.e.  we have
\begin{equation}\label{eq10} 
\delta\phi(x)=\frac{1}{\left(2\pi\right)^{3/2}}\int d^{3}\mathbf{k}\,\tilde{D}\left(\mathbf{k}\right)e^{i k_{1}\cdot x}+\frac{1}{(2\pi)^{3/2}}\int d^{3}\mathbf{k}\;\left(-\frac{1}{2}\right)\tilde{A}\left(\mathbf{k}\right)e^{ik_{2}\cdot x}+ c.c.\ ,
\end{equation}
and
\begin{equation}\label{eq11}
\delta\lambda(x)=\frac{1}{\left(2\pi\right)^{3/2}}\int d^{3}\mathbf{k}\,\left(-a_{1}\tilde{D}\left(\mathbf{k}\right)\right)e^{i k_{1}\cdot x}+\frac{1}{(2\pi)^{3/2}}\int d^{3}\mathbf{k}\;\left(\frac{a_{1}}{2}\right)\tilde{A}\left(\mathbf{k}\right)e^{ik_{2}\cdot x}+ c.c.\ .
\end{equation}
Now,  carrying out the spatial Fourier transform of Eq. \eqref{eqgcaseB} and by means of Eqs. \eqref{eq4}, \eqref{eq8} and \eqref{eq9}, we get 
\begin{equation}
6a_{1}\left(\partial_{0}^{2}+|\mathbf{k}|^{2}\right)\tilde{h}_{\mu\nu}\left(t,\mathbf{k}\right)+a_{1}\eta_{\mu\nu}k_{2}^{2}\tilde{A}\left(\mathbf{k}\right)e^{i\omega_{2}t}+2a_{1}\left(k_{2}\right)_{\mu}\left(k_{2}\right)_{\nu}\tilde{A}\left(\mathbf{k}\right)e^{i\omega_{2}t}=0\ ,
\end{equation}
that simplifying gives
\begin{equation}\label{eq20}
\left(\partial_{0}^{2}+|\mathbf{k}|^{2}\right)\tilde{h}_{\mu\nu}\left(t,\mathbf{k}\right)=-\left(\frac{\eta_{\mu\nu}k_{2}^{2}+2\left(k_{2}\right)_{\mu}\left(k_{2}\right)_{\nu}}{6}\right)\tilde{A}\left(\mathbf{k}\right)e^{i\omega_{2}t}\ .
\end{equation}
To solve the non-homogeneous Eq. \eqref{eq20}, first we get the related homogeneous solution linked to the massless wave $k_{1}^{2}=0$,
\begin{equation}
\tilde{h}_{\mu\nu}^{(o)}\left(t,\mathbf{k}\right)=C_{\mu\nu}^{(1)}\left(\mathbf{k}\right)e^{i\omega_{1}t}+C_{\mu\nu}^{(2)}\left(\mathbf{k}\right)e^{-i\omega_{1}t}\ ,
\end{equation}
and then a particular solution, linked to the massive wave $k_{2}^{2}=M^{2}$,  is   
\begin{equation}
\tilde{h}_{\mu\nu}^{(p)}\left(t,\mathbf{k}\right)=\frac{\eta_{\mu\nu}k_{2}^{2}+2\left(k_{2}\right)_{\mu}\left(k_{2}\right)_{\nu}}{6k_{2}^{2}}\tilde{A}\left(\mathbf{k}\right)e^{i\omega_{2}t}\ .
\end{equation}
Remembering that the solutions must be real, from the following decomposition of spatial coordinates, we have 
\begin{equation}
h_{\mu\nu}\left(x\right)=\frac{1}{\left(2\pi\right)^{3/2}}\int d^{3}\mathbf{k}\,\tilde{h}_{\mu\nu}\left(t,\mathbf{k}\right)e^{-i\mathbf{k}\cdot\mathbf{x}}\ ,
\end{equation}
and then we can reconstruct  $h_{\mu\nu}\left(x\right)$ as
\begin{equation}\boxed{
\begin{split}\label{GWNLGGost}
h_{\mu\nu}\left(x\right)=&\frac{1}{\left(2\pi\right)^{3/2}}\int d^{3}\mathbf{k}\,C_{\mu\nu}\left(\mathbf{k}\right)e^{i k_{1}\cdot x}\\+&\frac{1}{\left(2\pi\right)^{3/2}}\int d^{3}\mathbf{k}\left[\frac{1}{3}\left(\frac{\eta_{\mu\nu}}{2}+\frac{\left(k_{2}\right)_{\mu}\left(k_{2}\right)_{\nu}}{k_{2}^{2}}\right)\right]\tilde{A}\left(\mathbf{k}\right)e^{i k_{2}\cdot x}+ c.c.
\end{split}
}\ ,
\end{equation}
that is, the gravitational waves in non-local linear gravity.  For a similar approach in higher order gravity theories see \cite{CCC}. It is worth noticing that the part of solution due to non-locality,  related to $k_2^2\neq 0$, appears only if the constraint \eqref{caseB} is verified.

 Eq.\eqref{GWNLGGost} has the disadvantage of presenting ghost modes, so it is more useful to choose a suitable  gauge in order to suppress  waves without physical meaning.  In this perspective,  perturbing field Eqs. \eqref{eqfieldginmatter}, \eqref{eqfieldphi} and \eqref{eqfieldlambda} to first order in $h_{\mu\nu}$, $\delta\phi$ and $\delta\lambda$,  we get
\begin{equation}
\left(1+a_{1}\phi-\lambda\right)^{(0)}G_{\mu\nu}^{(1)}+\left(g_{\mu\nu}\Box-\nabla_{\mu}\nabla_{\nu}\right)^{(0)}\left(1+a_{1}\phi-\lambda\right)^{(1)}=\kappa^{2}T_{\mu\nu}^{(0)}\ ,
\end{equation}
that is 
\begin{equation}\label{eq30}
\left(1+a_{1}\phi_{0}-\lambda_{0}\right)G_{\mu\nu}^{(1)}+\left(\eta_{\mu\nu}\Box-\partial_{\mu}\partial_{\nu}\right)\left(a_{1}\delta\phi-\delta\lambda\right)=\kappa^{2}T_{\mu\nu}^{(0)}\ ,
\end{equation}
where $\Box=\eta^{\mu\nu}\partial_{\mu}\partial_{\nu}$, and
\begin{equation}\label{eq31}
\Box\delta\phi=R^{(1)}\ ,
\end{equation}
\begin{equation}\label{eq32}
\Box\delta\lambda=-a_{1}R^{(1)}\ .
\end{equation}
In a particular coordinates frame $\left\{x^{\mu}\right\}$, being our equations gauge invariant, we define a new gauge as 
\begin{equation}\label{newgauge}
\bar{h}_{\mu\nu}=h_{\mu\nu}-\frac{1}{2}\eta_{\mu\nu}h-\frac{\eta_{\mu\nu}}{1+a_{1}\phi_{0}-\lambda_{0}}\left(a_{1}\delta\phi-\delta\lambda\right)\ ,
\end{equation}
such that, in our reference frame, it is 
\begin{equation}
\partial_{\mu}\bar{h}^{\mu\nu}=0\ ,
\end{equation}
with $1+a_{1}\phi_{0}-\lambda_{0}\neq 0$.  The trace of Eq. \eqref{newgauge} yields
\begin{equation}
\bar{h}=-h-\frac{4}{1+a_{1}\phi_{0}-\lambda_{0}}\left(a_{1}\delta\phi-\delta\lambda\right)	\ .
\end{equation} 
In this gauge, we get the following expressions to first order for the Ricci tensor $R_{\mu\nu}$,  the Ricci scalar $R$ and the Einstein tensor $G_{\mu\nu}$
\begin{equation}
R_{\mu\nu}^{(1)}=-\frac{1}{2}\Box h_{\mu\nu}+\frac{1}{1+a_{1}\phi_{0}-\lambda_{0}}\partial_{\mu}\partial_{\nu}\left(a_{1}\delta\phi-\delta\lambda\right)\ ,
\end{equation}
\begin{equation}
R^{(1)}=-\frac{1}{2}\Box h+\frac{1}{1+a_{1}\phi_{0}-\lambda_{0}}\Box\left(a_{1}\delta\phi-\delta\lambda\right)\ ,
\end{equation}
\begin{equation}
G_{\mu\nu}^{(1)}=-\frac{1}{2}\Box h_{\mu\nu}+\frac{1}{4}\eta_{\mu\nu}\Box h+\frac{1}{1+a_{1}\phi_{0}-\lambda_{0}}\left(\partial_{\mu}\partial_{\nu}-\frac{1}{2}\eta_{\mu\nu}\Box\right)\left(a_{1}\delta\phi-\delta\lambda\right)\ .
\end{equation}
In terms of barred quantities $\bar{h}_{\mu\nu}$ and $\bar{h}$,  we obtain 
\begin{equation}
R^{(1)}=\frac{1}{2}\Box\bar{h}+\frac{3}{1+a_{1}\phi_{0}-\lambda_{0}}\Box\left(a_{1}\delta\phi-\delta\lambda\right)\ ,
\end{equation}
\begin{equation}
G_{\mu\nu}^{(1)}=-\frac{1}{2}\Box\bar{h}_{\mu\nu}-\frac{1}{1+a_{1}\phi_{0}-\lambda_{0}}\left(\eta_{\mu\nu}\Box-\partial_{\mu}\partial_{\nu}\right)\left(a_{1}\delta\phi-\delta\lambda\right)\ ,
\end{equation}
which replaced in Eqs. \eqref{eq30}, \eqref{eq31} and \eqref{eq32} give, in vacuum,
\begin{equation}\label{eq35}
\Box\bar{h}_{\mu\nu}=0\ ,
\end{equation}
\begin{equation}\label{eq36}
\Box\delta\phi=\frac{3}{1+a_{1}\phi_{0}-\lambda_{0}}\left(a_{1}\Box\delta\phi-\Box\delta\lambda\right)\ ,
\end{equation}
and
\begin{equation}\label{eq37}
\Box\delta\lambda=-\frac{3a_{1}}{1+a_{1}\phi_{0}-\lambda_{0}}\left(a_{1}\Box\delta\phi-\Box\delta\lambda\right)\ ,
\end{equation}
because  Eq. \eqref{eq35} implies $\Box\bar{h}=0$. Eqs. \eqref{eq36} and \eqref{eq37} can be rewritten as
\begin{equation}\label{eq38a}
\left(\frac{1+a_{1}\phi_{0}-\lambda_{0}}{3}-a_{1}\right)\Box\delta\phi+\Box\delta\lambda=0\ ,
\end{equation}
and 
\begin{equation}\label{eq38b}
a_{1}\Box\delta\phi+\left(\frac{1+a_{1}\phi_{0}-\lambda_{0}}{3a_{1}}-1\right)\Box\delta\lambda=0\ .
\end{equation} 
The homogeneous linear system  \eqref{eq38a} and \eqref{eq38b}, in $\Box\delta\phi$ and $\Box\delta\lambda$ variables, admits trivial and non-trivial solutions depending on  the determinant of the following matrix 
\begin{equation}
 \text{det}\begin{pmatrix} 
    \frac{1+a_{1}\phi_{0}-\lambda_{0}}{3}-a_{1}& 1  \\ 
     a_{1} & \frac{1+a_{1}\phi_{0}-\lambda_{0}}{3a_{1}}-1 \\ 
  \end{pmatrix}=\frac{1}{a_{1}}\left(1+a_{1}\phi_{0}-\lambda_{0}-6a_{1}\right)\left(1+a_{1}\phi_{0}-\lambda_{0}\right)\ ,
\end{equation}
which can be equal to zero or non-equal to zero,  depending on   $1+a_{1}\phi_{0}-\lambda_{0}\neq 0$. So, if the determinant does not vanish, under the constraint 
\begin{equation}\label{constraint1}
1+a_{1}\phi_{0}-\lambda_{0}\neq6a_{1}\ ,
\end{equation}
we get  the trivial solution
\begin{equation}
\Box\delta\phi=\Box\delta\lambda=0\ ,
\end{equation} 
from which, it  follows 
\begin{equation}\label{39}
\Box \left(a_{1}\delta\phi-\delta\lambda\right)=0\ .
\end{equation}
A particular solution of  Eq. \eqref{39} is
\begin{equation}
a_{1}\delta\phi\left(x\right)=\delta\lambda\left(x\right)\,.
\end{equation}
Putting this solution together with Eq. \eqref{eq35} and Eq. \eqref{newgauge},  we get   massless transverse gravitational waves with helicity 2,  associated to $k_{1}^{2}=0$ under the constraint \eqref{constraint1}
\begin{equation}\label{masslesswave}\boxed{
h_{\mu\nu}^{(k_{1})}\left(x\right)=\frac{1}{\left(2\pi\right)^{3/2}}\int d^{3}\mathbf{k}\,\tilde{C}_{\mu\nu}\left(\mathbf{k}\right)e^{i k_{1}\cdot x}+c.c.
}\ ,
\end{equation}
where $\tilde{C}_{\mu\nu}$ is the traceless polarization tensor, i.e. $\tilde{C}^{\sigma}_{\phantom{\sigma}\sigma}=0$,  in a suitable gauge that leaves $\partial_{\mu}\bar{h}^{\mu\nu}=0$ invariant, as in the case \eqref{caseA}.

The non-trivial solution is obtained by imposing  that the determinant of the  above matrix  is equal to zero,  which means 
\begin{equation}\label{constraint2}
1+a_{1}\phi_{0}-\lambda_{0}=6a_{1}\ ,
\end{equation}
that gives again 
\begin{equation}\label{constraint_1}
a_{1}\Box\delta\phi-\Box\delta\lambda\neq 0\ ,
\end{equation}
or equivalently 
\begin{equation}
\Box h\neq 0\ ,
\end{equation}
as in the case \eqref{caseB}.
Then, from the inverse gauge under the constraint \eqref{constraint2}
\begin{equation}\label{eq40}
h_{\mu\nu}=\bar{h}_{\mu\nu}-\frac{1}{2}\eta_{\mu\nu}\bar{h}-\frac{\eta_{\mu\nu}}{6a_{1}}\left(a_{1}\delta\phi-\delta\lambda\right)\ ,
\end{equation}
we choose the non-homogeneous wave equation
\begin{equation}\label{eq41}
\frac{2}{3}\Box\left(a_{1}\delta\phi-\delta\lambda\right)=-a_{1}\Box h=-a_{1}g\left(x\right)\ ,
\end{equation}
with $g\left(x\right)\in L(\mathbf{R}^{4})$, as already defined in \eqref{possiblefunctiong}. The particular solution of Eq. \eqref{eq41} is
\begin{equation}\label{eq42}
a_{1}\delta\phi(x)-\delta\lambda(x)=\frac{1}{(2\pi)^{3/2}}\int d^{3}\mathbf{k}\;\left(-\frac{3}{2}a_{1}\right)\tilde{A}\left(\mathbf{k}\right)e^{ik_{2}\cdot x}+ c.c.\ ,
\end{equation}
with any $k_{2}^{2}=M^{2}$. Inserting Eq. \eqref{eq42} into Eq. \eqref{eq40},  we obtain a massive wave 
\begin{equation}\label{massivewave}\boxed{
h_{\mu\nu}^{(k_{2})}\left(x\right)=\frac{1}{(2\pi)^{3/2}}\int d^{3}\mathbf{k}\;\frac{\eta_{\mu\nu}}{4}\tilde{A}\left(\mathbf{k}\right)e^{ik_{2}\cdot x}+ c.c.
}\ .
\end{equation}
In general under the constraint \eqref{constraint2},  the gravitational waves for non-local  $f(R,\Box^{-1}R)$ gravity are both massless and massive waves,  that is 
\begin{equation}\boxed{
\begin{split}
h_{\mu\nu}\left(x\right)=&\frac{1}{\left(2\pi\right)^{3/2}}\int d^{3}\mathbf{k}\,\tilde{C}_{\mu\nu}\left(\mathbf{k}\right)e^{i k_{1}\cdot x}\\
+&\frac{1}{(2\pi)^{3/2}}\int d^{3}\mathbf{k}\;\frac{\eta_{\mu\nu}}{4}\tilde{A}\left(\mathbf{k}\right)e^{ik_{2}\cdot x}+ c.c.
\end{split}
}\ .
\end{equation}
Also here the solution component,  related to $k_2^2\neq 0$,  appears if  the constraint \eqref{constraint2} is satisfied.
With these results in mind, let us investigate gravitational wave polarizations in non-local gravity.

\section{Polarizations via geodesic deviation }\label{c}
A useful tool to study the polarization of gravitational radiation is the use of geodetic deviation produced by the wave when it invests a small region of spacetime, as the relative acceleration measured between nearby geodesics.  Hence, we start from a wave $h_{\mu\nu}\left(t-v_{g}z\right)$  propagating in  $+\hat{z}$ direction in a local proper reference frame,  where $v_{g}$ is the group velocity in units where $c=1$ defined as
\begin{equation}
v_{g}=\frac{d\omega}{d k_{z}}=\frac{c^{2}k_{z}}{\omega}\ ,
\end{equation} 
and let us consider the equation for geodesic deviation 
\begin{equation}\label{eqdevgeoelectric}
\ddot x^{i}=-R^{i}_{\phantom{i}0k0}x^{k}\ ,
\end{equation}
where the Latin index range over the set $\left\{1,2,3\right\}$ and $R^{i}_{\phantom{i}0k0}$ are the only measurable components called the electric ones \cite{STRAGR}.  After replacing the linearized electric components of the Riemann tensor $R^{\left(1\right)}_{\phantom{1}i0j0}$,  expressed in terms of the metric perturbation $h_{\mu\nu}$ 
\begin{equation}
R^{\left(1\right)}_{\phantom{1}i0j0}=\frac{1}{2}\left(h_{i0,j0}+h_{j0,i0}-h_{ij,00}-h_{00,ij}\right)\ ,
\end{equation}
in Eq.\eqref{eqdevgeoelectric},  we get a linear non-homogeneus system of differential equations  
\begin{equation}\label{eqdevgeolinear}
\begin{cases}
\ddot x(t)=-\frac{1}{2}h_{11,00}\;x-\frac{1}{2}h_{12,00}\;y+\frac{1}{2}\left(h_{10,03}-h_{13,00}\right)z\\
\ddot y(t)=-\frac{1}{2}h_{12,00}\;x-\frac{1}{2}h_{22,00}\;y+\frac{1}{2}\left(h_{02,03}-h_{23,00}\right)z\\
\ddot z(t)=\frac{1}{2}\left(h_{01,03}-h_{13,00}\right)x+\frac{1}{2}\left(h_{02,03}-h_{23,00}\right)y+\frac{1}{2}\left(2h_{03,03}-h_{33,00}-h_{00,33}\right)z\,.
\end{cases}
\end{equation}
For a massless plane wave travelling in  $+\hat{z}$ direction,  that is $k_{1}^{2}=0$,  which propagates at speed $c$,  if we keep $\mathbf{k}$ fixed and $k_{1}^{\mu}=\left(\omega_{1},0,0,k_{z}\right)$,    Eq.\eqref{masslesswave} yields    
\begin{equation}\label{FirstOrderTetradmassless}
h^{(k_{1})}_{\mu\nu}\left(t,z\right)=\sqrt{2}\left[\tilde{\epsilon}^{(+)}\left(\omega_{1}\right)\epsilon^{(+)}_{\mu\nu}+\tilde{\epsilon}^{(\times)}\left(\omega_{1}\right)\epsilon^{(\times)}_{\mu\nu}\right]e^{i\omega_{1}\left(t-z\right)}+c.c.\ ,
\end{equation}
where 
\begin{equation}
\epsilon^{(+)}_{\mu\nu}=\frac{1}{\sqrt{2}}
\begin{pmatrix} 
0 & 0 & 0 & 0 \\
0 & 1 & 0 & 0 \\
0 & 0 & -1 & 0 \\
0 & 0 & 0 & 0
\end{pmatrix}\ ,
\end{equation}
\begin{equation}
\epsilon^{(\times)}_{\mu\nu}=\frac{1}{\sqrt{2}}
\begin{pmatrix} 
0 & 0 & 0 & 0 \\
0 & 0 & 1 & 0 \\
0 & 1 & 0 & 0 \\
0 & 0 & 0 & 0
\end{pmatrix}\ ,
\end{equation}
and $\omega_{1}=k_{z}$.  Furthermore,  for a massive plane wave propagating in $+\hat{z}$ direction,  that is $k_{2}^{2}\neq 0$,  always keeping $\mathbf{k}$ fixed and instead with $k_{2}^{\mu}=\left(\omega_{2},0,0,k_{z}\right)$,   Eq.\eqref{massivewave} becomes 
\begin{equation}\label{FirstOrderTetradmassive}
h^{(k_{2})}_{\mu\nu}\left(t,z\right)=\frac{\tilde{A}\left(k_{z}\right)}{4}\eta_{\mu\nu}e^{i\left(\omega_{2}t-k_{z}z\right)}+c.c.\ ,
\end{equation}
where here the propagation speed is less than $c$.  In a more compact form,  the metric linear perturbation  $h_{\mu\nu}$,  travelling in the $+\hat{z}$ direction, assuming  $\mathbf{k}$ fixed,  is
\begin{equation} 
h_{\mu\nu}\left(t,z\right)=\sqrt{2}\left[\tilde{\epsilon}^{(+)}\left(\omega_{1}\right)\epsilon^{(+)}_{\mu\nu}+\tilde{\epsilon}^{(\times)}\left(\omega_{1}\right)\epsilon^{(\times)}_{\mu\nu}\right]e^{i\omega_{1}\left(t-z\right)}+\tilde{\epsilon}^{\left(s\right)}_{\mu\nu}\left(k_{z}\right)e^{i\left(\omega_{2}t-k_{z}z\right)}+c.c.\ ,
\end{equation}
where $\tilde{\epsilon}^{\left(s\right)}_{\mu\nu}$ is the polarization tensor associated to the mixed scalar mode 
\begin{equation}
\tilde{\epsilon}^{\left(s\right)}_{\mu\nu}\left(k_{z}\right)=\left(\epsilon^{(TT)}_{\mu\nu}-\sqrt{2}\epsilon_{\mu\nu}^{(b)}-\epsilon_{\mu\nu}^{(l)}\right)\frac{\tilde{A}\left(k_{z}\right)}{4}\ .
\end{equation}
and the polarization tensors are explicitly given by 
\begin{align}
\epsilon^{(TT)}_{\mu\nu}&=
\begin{pmatrix} 
1 & 0 & 0 & 0 \\
0 & 0 & 0 & 0 \\
0 & 0 & 0 & 0 \\
0 & 0 & 0 & 0
\end{pmatrix}\ , &
\epsilon^{(b)}_{\mu\nu}&=\frac{1}{\sqrt{2}}
\begin{pmatrix} 
0 & 0 & 0 & 0 \\
0 & 1 & 0 & 0 \\
0 & 0 & 1 & 0 \\
0 & 0 & 0 & 0
\end{pmatrix}\ , &
\epsilon^{(l)}_{\mu\nu}&=
\begin{pmatrix} 
0 & 0 & 0 & 0 \\
0 & 0 & 0 & 0 \\
0 & 0 & 0 & 0 \\
0 & 0 & 0 & 1
\end{pmatrix}\ .
\end{align}
where the set of polarization tensors $\left\{\epsilon_{\mu\nu}^{\left(+\right)}, \epsilon_{\mu\nu}^{\left(\times\right)}, \epsilon_{\mu\nu}^{\left(TT\right)}, \epsilon_{\mu\nu}^{\left(b\right)}, \epsilon_{\mu\nu}^{\left(l\right)}\right\}$ satisfy the orthonormality relations
\begin{equation}
\text{Tr}\left\{\epsilon^{(a)}\epsilon^{(b)}\right\}=\epsilon_{\mu\nu}^{(a)}\epsilon^{(b)\mu\nu}=\delta^{a,b}\quad\text{with}\quad a,b\in\left\{+, \times, TT, b, l\right\}\ .
\end{equation} 
The scalar mode $\tilde{\epsilon}^{\left(s\right)}_{\mu\nu}$ is a mixed state obtained from a combination of longitudinal and transverse scalar modes which is produced by the single degree of freedom $\tilde{A}$ as for $f(R)$ gravity which has three d.o.f.: $\tilde{\epsilon}^{(+)}$, $\tilde{\epsilon}^{(\times)}$ and $\tilde{A}$ \cite{LGHL, IKEDA}.  However, as we will see, the transverse component weighs more than the longitudinal one.  In fact,  the polarization tensor $\tilde{\epsilon}^{\left(s\right)}_{\mu\nu}$,  restricted to  spatial components $\tilde{\epsilon}^{\left(s\right)}_{i,j}$, is provided by 
\begin{equation}
\tilde{\epsilon}^{\left(s\right)}_{i,j}=\left(\sqrt{2}\epsilon_{i,j}^{(b)}+\epsilon_{i,j}^{(l)}\right)\frac{\tilde{A}\left(k_{z}\right)}{4}\ ,
\end{equation}
where $(i,j)$ range over $(1,2,3)$.  Hence,  from Eqs.\eqref{eqdevgeolinear} in the case of massless plane waves $h_{\mu\nu}^{(k_{1})}$,  we have 
\begin{equation}
\begin{cases}
\ddot x(t)=\frac{1}{2}\omega_{1}^{2}\left[\tilde{\epsilon}^{\left(+\right)}\left(\omega_{1}\right)x+\tilde{\epsilon}^{\left(\times\right)}\left(\omega_{1}\right)y\right]e^{i\omega_{1}\left(t-z\right)}+c.c. \\ \\
\ddot y(t)=\frac{1}{2}\omega_{1}^{2}\left[\tilde{\epsilon}^{\left(\times\right)}\left(\omega_{1}\right)x-\tilde{\epsilon}^{\left(+\right)}\left(\omega_{1}\right)y\right]e^{i\omega_{1}\left(t-z\right)}+c.c.\\ \\
\ddot z(t)=0
\end{cases}\ ,
\end{equation}
that give us the two standard transverse tensor polarizations predicted by General Relativity,  conventionally called plus and cross modes.   \\

Otherwise,  in the case of a massive plane wave $h^{(k_{2})}_{\mu\nu}$ with $M^2=\omega^{2}_{2}-k^{2}_{z}$,  the linearized geodesic deviation  Eq.\eqref{eqdevgeolinear} becomes
\begin{equation}
\begin{cases}
\ddot x(t)=-\frac{1}{8}\omega^{2}_{2}\tilde{A}\left(k_{z}\right)x e^{i\left(\omega_{2}t-k_{z}z\right)}+c.c.\\ \\
\ddot y(t)=-\frac{1}{8}\omega^{2}_{2}\tilde{A}\left(k_{z}\right)y e^{i\left(\omega_{2}t-k_{z}z\right)}+c.c.\\ \\
\ddot z(t)=-\frac{1}{8}M^{2}\tilde{A}\left(k_{z}\right)ze^{i\left(\omega_{2}t-k_{z}z\right)}+c.c.
\end{cases}\ ,
\end{equation}
that can be integrated, assuming that $h_{\mu\nu}\left(t,z\right)$ is small,  as 
\begin{equation}	\label{solutionsldgs}
\begin{cases}
x(t)=x(0)+\frac{1}{8}\tilde{A}\left(k_{z}\right)x(0)e^{i\left(\omega_{2}t-k_{z}z\right)}+c.c.\\ \\
y(t)=y(0)+\frac{1}{8}\tilde{A}\left(k_{z}\right)y(0)e^{i\left(\omega_{2}t-k_{z}z\right)}+c.c.\\ \\
z(t)=z(0)+\frac{1}{8\omega^{2}_{2}}M^{2}\tilde{A}\left(k_{z}\right)z(0)e^{i\left(\omega_{2}t-k_{z}z\right)}+c.c.
\end{cases}\ .
\end{equation}
If we suppose that $M^2$ is very small, which happens when we are sufficiently far from the  radiation source,  and that becomes zero for exactly null plane waves,  $k^{2}=0$,  then we can expand our results with respect to a parameter $\gamma$ defined as \cite{CMW} 
\begin{equation}
\gamma=\left(\frac{c}{v_{g}}\right)^{2}-1\ ,
\end{equation} 
which takes into account the difference in speed between nearly null waves with speed $v_{g}$ and null ones with speed $c$. Thus,  keeping $k_{z}$ fixed and by using Landau symbols,  namely little-$o$ and big-$\mathcal{O}$ notation,  we can expand in terms of our parameter $\gamma$ the following quantities 
\begin{equation}
\frac{M^{2}}{k_{z}^{2}}=\frac{1}{k_{z}^{2}}\left[\frac{\omega^{2}_{2}}{c^{2}}-k_{z}^{2}\right]=\left(\frac{\omega_{2}}{c k_{z}}\right)^{2}-1=\gamma\ ,
\end{equation}
and in units where $c=1$
\begin{equation}
\frac{\omega_{2}}{k_{z}}=\sqrt{1+\gamma}=\left(1+\frac{1}{2}\gamma\right)+o\left(\gamma\right)\ ,
\end{equation}
\begin{equation}
e^{i\left(\omega_{2}t-k_{z}z\right)}=e^{ik_{z}\left(t-z\right)}+\mathcal{O}\left(\gamma\right)\ ,
\end{equation}
that implies
\begin{equation}
\frac{M^{2}}{\omega_{2}^{2}}=\gamma+o\left(\gamma\right)\ .
\end{equation}
Within the first order in $\gamma$,  the solution   \eqref{solutionsldgs} for the mode $\omega_{2}$ give us 
\begin{equation}	
\begin{cases}
x(t)=x(0)+\frac{1}{8}\tilde{A}\left(k_{z}\right)x(0)e^{i k_{z}\left(t-z\right)}+\mathcal{O}\left(\gamma\right)+c.c.\\ \\
y(t)=y(0)+\frac{1}{8}\tilde{A}\left(k_{z}\right)y(0)e^{i k_{z}\left(t-z\right)}+\mathcal{O}\left(\gamma\right)+c.c.\\ \\
z(t)=z(0)+\frac{1}{8}\gamma\tilde{A}\left(k_{z}\right)z(0)e^{i k_{z}\left(t-z\right)}+\mathcal{O}\left(\gamma^{2}\right)+c.c.
\end{cases}\ ,
\end{equation}
that is 
\begin{equation}
\Delta z(t)=o\left(\Delta x(t)\right)\quad \text{for}\quad\gamma\to 0\ ,
\end{equation}
and then only the breathing tensor polarization $\epsilon^{(b)}$ survives because longitudinal modes are infinitesimal in higher order than transverse modes when $\gamma$ tends to zero.  Thus when a GW strikes a sphere of particles of radius $r=\sqrt{x^2(0)+y^{2}(0)+z^{2}(0)}$, this  will be distorted into an ellipsoid described by
\begin{equation}
\left(\frac{x}{\rho_{1}(t)}\right)^{2}+\left(\frac{y}{\rho_{1}(t)}\right)^{2}+\left(\frac{z}{\rho_{2}(t)}\right)^{2}=r^{2}\ ,
\end{equation}
where $\rho_{1}(t)=1+\frac{1}{4}\tilde{A}\left(k_{z}\right)\cos\left[k_{z}\left(t-z\right)+\phi\right]+\mathcal{O}\left(\gamma\right)$ and $\rho_{2}(t)=1+\mathcal{O}\left(\gamma\right)$, that, at zero order in $\gamma$ only $\rho_{1}$,  is varying between their maximum and minimum values. This ellipsoid swings only on $xy$-plane between two circumferences of minimum and maximum radius and represents an additional transverse scalar polarization which has zero helicity  within the lowest order in $\gamma$ \cite{PW}. 

According to these considerations, the d.o.f. of non-local  $f(R,\Box^{-1}R)$ gravity are three: two of these,  $\tilde{\epsilon}^{\left(+\right)}$ and $\tilde{\epsilon}^{\left(\times\right)}$, give rise to the standard   tensor modes of General Relativity while the degree of freedom $\tilde{A}$ generates a further breathing scalar mode.  In summary,   $f(R,\Box^{-1}R)$ gravity has three polarizations,  namely two massless 2-helicity  tensor modes and one massive 0-helicity scalar mode,  all purely transverse within the lowest order in $\gamma$,  exactly like $f(R)$ gravity (see for a discussion \cite{IKEDA, AMA, MY, BCLN, CCL, CCDLV, KPJ}). 

\section{Polarizations via Newman-Penrose  formalism}\label{d}
A further approach to study polarizations can be obtained by adopting  the Newman-Penrose (NP) formalism  for low-mass gravitational waves \cite{CMW}. Even if it is not directly applicable to  massive waves because it was, in origin,   worked out  for  massless waves, it is possible  to  generalize it to low-mass waves  propagating along nearly null geodesics \cite{NP}. It is worth noticing that the  little group $E\left(2\right)$ classification fails for massive waves but can be recovered within the first order in $\gamma$. 

Let us introduce a new basis, namely a local quasi-orthonormal null tetrad basis $\left(k, l, m, \bar{m}\right)$ defined as \cite{PW, ELL, ELLWW}
\begin{align}
k&=\partial_{t}+\partial_{z}\ ,
& l&=\frac{1}{2}\left(\partial_{t}-\partial_{z}\right)\ ,\\
m&=\frac{1}{\sqrt{2}}\left(\partial_{x}+i\partial_{y}\right)\ ,
& \bar{m}&=\frac{1}{\sqrt{2}}\left(\partial_{x}-i\partial_{y}\right)\ ,
\end{align}
which,  adopting the Minkowski metric tensor $\eta_{\mu\nu}$ of signature $-2$,  satisfies the relations
\begin{gather}
k\cdot l=-m\cdot\bar{m}=1\ , \nonumber\\
k\cdot k=l\cdot l=m\cdot m=\bar{m}\cdot\bar{m}=0\ ,\\
k\cdot m=k\cdot\bar{m}=l\cdot m=l\cdot\bar{m}=0\ ,\nonumber
\end{gather}
that is
\begin{equation}
\eta^{\mu\nu}=2k^{(\mu}l^{\nu)}-2m^{(\mu}\bar{m}^{\nu)}\ .
\end{equation}
Therefore we can raise and lower the tetrad indices by the metric of the tetrad $\eta_{ab}$
\begin{equation}
\eta_{ab}=\eta^{ab}=
\begin{pmatrix}
0	&	1	&	0	&	0\\
1	&	0	&	0	&	0\\
0	&	0	&	0	&	-1\\
0	&	0	&	-1	&	0\\
\end{pmatrix} \ ,
\end{equation}
where $(a,b)$ run over $(k, l, m, \bar{m})$. We now split the Riemann tensor into three irreducible parts, namely:  Weyl tensor, traceless Ricci tensor and Ricci scalar, known as Newmann-Penrose quantities. The four-dimensional Weyl tensor $C_{\mu\nu\rho\sigma}$ is defined as
\begin{equation}
C_{\mu\nu\rho\sigma}=R_{\mu\nu\rho\sigma}-2g_{[\mu|[\rho}R_{\sigma]|\nu]}+\frac{1}{3}g_{\mu[\rho}g_{\sigma]\nu}R\ ,
\end{equation}
and in tetrad form becomes
 \begin{equation}\label{TetradWeylTensor}
 C_{abcd}=R_{abcd}-2\eta_{[a|[c}R_{d]|b]}+\frac{1}{3}\eta_{a[c}\eta_{d]b}R\ ,
 \end{equation}
taking into account that the tetrad components of the generic  tensor $P_{a b c d \dots}$ express in terms of the local coordinate basis are 
\begin{equation}
P_{a b c d \dots}=P_{\alpha,\beta\gamma\delta\dots}a^{\alpha}b^{\beta}c^{\gamma}d^{\delta}\cdots
\end{equation}
where $(a, b, c, d, \dots)$ run over $(k, l, m, \bar{m})$. 
The fifteen NP-amplitudes are specifically the five complex Weyl-NP $\Psi$  scalars, expressed in tetrad components of the Weyl tensor as
\begin{eqnarray}
 \Psi_{0}\equiv -C_{kmkm}\ ,\nonumber\\
\Psi_{1}\equiv -C_{klkm}\ ,\nonumber\\
 \Psi_{2}\equiv -C_{km\bar{m}l}\ ,\\
 \Psi_{3}\equiv -C_{kl\bar{m}l}\ ,\nonumber\\
 \Psi_{4}\equiv -C_{\bar{m}l\bar{m}l}\ ,\nonumber
\end{eqnarray}
 and the ten Ricci-NP scalars $\Phi$, $\Lambda$, expressed in tetrad components of  Ricci tensor  as
\begin{eqnarray*}
\Phi_{02}\equiv-\frac{1}{2}R_{mm\ ,}\\
\end{eqnarray*} 
\begin{equation*}
\begin{cases}
\Phi_{01}\equiv-\frac{1}{2}R_{km}\\
\Phi_{12}\equiv-\frac{1}{2}R_{lm}\\
\end{cases}\ ,
\end{equation*}
\begin{equation*}
\begin{cases}
\Phi_{00}\equiv-\frac{1}{2}R_{kk}\\
\Phi_{11}\equiv-\frac{1}{4}\left(R_{kl}+R_{m\bar{m}}\right)\\
\Phi_{22}\equiv-\frac{1}{2}R_{ll}
\end{cases}\ ,
\end{equation*} 
\begin{equation*}
\begin{cases}
\Phi_{10}\equiv-\frac{1}{2}R_{k\bar{m}}=\Phi_{01}^{*}\\
\Phi_{21}\equiv-\frac{1}{2}R_{l\bar{m}}=\Phi_{21}^{*}\\
\end{cases}\ ,
\end{equation*} 
\begin{eqnarray}
 \Phi_{20}\equiv-\frac{1}{2}R_{\bar{m}\bar{m}}=\Phi_{02}^{*}\ ,\nonumber\\
\Lambda=\frac{R}{24}\ .
\end{eqnarray} 
In order to expand the low-mass gravitational radiation in terms of null plane waves, we first define the "wave"  four-vector $\left(k_{2}^{\prime}\right)^{\mu}$ in units where $c=1$ associated to the nearly null plane wave propagating in positive $z$ direction as 
\begin{equation}
k_{2}^{\mu}=\left(\omega_{2}, 0, 0, k_{z}\right)=\omega_{2}k_{2}^{\prime\mu}\ ,
\end{equation}
with
\begin{equation}
k^{\prime\mu}_{2}=\left(1, 0, 0, v_{g}\right)\ ,
\end{equation}
and we set the time retarded $\tilde{u}$ as
\begin{equation}
\tilde{u}=t-v_{g}z\ ,
\end{equation}
that gives us
\begin{equation}
\nabla_{\mu}\tilde{u}=\left(k_{2}^{\prime}\right)_{\mu}\ .
\end{equation}
Now,  we expand $k_{2}^{\prime}$ with respect to our null tetrads basis 
\begin{equation}\label{kprimeresklm}
k_{2}^{\prime\mu}=\left(1+\gamma_{k}\right)k^{\mu}+\gamma_{l}l^{\mu}+\gamma_{m}m^{\mu}+\bar{\gamma}_{m}\bar{m}^{\mu}\ ,
\end{equation} 
where the expansion coefficients $\gamma_{k}, \gamma_{l}, \gamma_{m}$ are of same order of the $\gamma$ of previous section.  Given the arbitrariness of the observer to orient its  reference system,  it is possible to perform orientation in such a way that  $k_{2}^{0}=k^{0}$ and $k_{2}^{3}\propto k^{3}$,  where $k^{0}$ and $k^{3}$ are the angular frequency and the third component of vector wave of its null wave,  respectively.  Therefore we obtain, from Eq. \eqref{kprimeresklm} $,\gamma_{l}=-2\gamma_{k}$ and $\gamma_{m}=0$ that yields  
\begin{equation}
k_{2}^{\prime\mu}=k^{\mu}+\gamma_{k}\left(k^{\mu}-2l^{\mu}\right)\ ,
\end{equation}
or 
\begin{equation}
k_{2}^{\prime\mu}=\left(1, 0, 0, 1+2\gamma_{k}\right)\ ,
\end{equation}
and, as already observed according to 
\begin{equation}
\gamma_{k}=-\frac{1}{4}\gamma+\small{o}\left(\gamma\right)\ ,
\end{equation}
the parameters $\gamma$ and $\gamma_{k}$ are of the same order.  The derivatives of Riemann tensor can be expressed as
\begin{equation}
R_{\alpha\beta\gamma\delta,\mu}=\frac{\partial R_{\alpha\beta\gamma\delta}}{\partial\tilde{u}}\nabla_{\mu}\tilde{u}=\left(k_{2}^{\prime}\right)_{\mu}\dot R_{\alpha\beta\gamma\delta}\ ,
\end{equation}
where the superscripted dot means the derative with respect to $\tilde{u}$. 
The  following identities 
\begin{equation}
R_{\alpha\beta\gamma\delta,k}=-2\gamma_{k}\dot R_{\alpha\beta\gamma\delta}\ ,
\end{equation}
\begin{equation}
R_{\alpha\beta\gamma\delta,l}=\left(1+\gamma_{k}\right)\dot R_{\alpha\beta\gamma\delta}\ ,
\end{equation}
\begin{equation}
R_{\alpha\beta\gamma\delta,m}=0\ ,
\end{equation}
combined with differential Bianchi identity 
\begin{equation}
R_{ab[cd,e]}=0\ ,
\end{equation}
involve that the only non-zero tetrad components of Riemann tensor $R_{\alpha\beta\gamma\delta}$ to zero order in  $\gamma_{k}$ are terms of form $R_{lplq}$ with $(p,q)$ range over $(k, m, \bar{m)}$.  In the nearly null plane waves framework,  only four complex NP tetrad components are independent and non-vanishing within the first order in $\gamma$, that is, from Eq. \eqref{TetradWeylTensor}, they are
\begin{align}
\Psi_{2}\left(\tilde{u}\right)&=-\frac{1}{6}R_{lklk}+\mathcal{O}\left(\gamma_{k}\right)\ ,\\
\Psi_{3}\left(\tilde{u}\right)&=-\frac{1}{2}R_{lkl\bar{m}}+\mathcal{O}\left(\gamma_{k}\right)\ ,\\
\Psi_{4}\left(\tilde{u}\right)&=-R_{l\bar{m}l\bar{m}}\ ,\\
\Phi_{22}\left(\tilde{u}\right)&=-R_{lml\bar{m}}\ .\\
\end{align} 
In terms of tetrad components of metric perturbation $h_{ab}$, they become 
\begin{align}
\Psi_{2}\left(\tilde{u}\right)&=\frac{1}{12}\ddot{h}_{kk}+\mathcal{O}\left(\gamma_{k}\right)\ ,\\
\Psi_{3}\left(\tilde{u}\right)&=\frac{1}{4}\ddot{h}_{k\bar{m}}+\mathcal{O}\left(\gamma_{k}\right)\ ,\\
\Psi_{4}\left(\tilde{u}\right)&=\frac{1}{2}\ddot{h}_{\bar{m}\bar{m}}+\mathcal{O}\left(\gamma_{k}\right)\ ,\\
\Phi_{22}\left(\tilde{u}\right)&=\frac{1}{2}\ddot{h}_{m\bar{m}}+\mathcal{O}\left(\gamma_{k}\right)\ .\\
\end{align}
These  real amplitudes, under the subgroup of Lorentz transformations which leaves  $\mathbf{k}_{2}$ unchanged, namely the little group $E(2)$, show the following  helicity values $s$:
\begin{align}
\Psi_{2}\quad s=0\ ,& &\Psi_{4\quad}  s=2\ ,&\\
\Psi_{3}\quad  s=1\ ,& &\Phi_{22}\quad  s=0\ .&
\end{align}
They allow the Petrov classification for a set of quasi-Lorentz invariant  GWs.  Remembering that our gravitational radiation  travels along the positive $+\hat{z}$ axis, we obtain the following four NP amplitudes expressed both in terms of the electric components of the Riemann tensor $R_{i0j0}$ and its linearized components \cite{BEMI, AMdeA, dePMM, WAG}. That is,  taking into account the identities 
\begin{align}
R_{kml\bar{m}}&=0&\quad&\rightarrow&\quad &R_{1313}+R_{2323}=R_{0101}+R_{0202}\\
R_{kmk\bar{m}}&=0&\quad&\rightarrow&\quad &-2R_{0131}-2R_{0232}=2\left(R_{0101}+R_{0202}\right)\\
R_{klkm}&=0&\quad&\rightarrow&\quad &R_{0301}=-R_{0331}\quad\text{and}\quad R_{0302}=-R_{0332}\\
R_{kmlm}&=0&\quad&\rightarrow&\quad &R_{3132}=R_{0102}\quad\text{and}\quad R_{3131}-R_{3232}=R_{0101}-R_{0202}\\
R_{kmkm}&=0&\quad&\rightarrow&\quad 
&\begin{cases}2R_{3101}-2R_{3202}=-R_{3131}+R_{3232}-R_{0101}+R_{0202}\\
R_{0102}=-R_{0132}-R_{0231}-R_{3132}\,,
\end{cases}
\end{align}
we get, in terms of $R_{i0j0}$, 
\begin{align}
\Psi_{2}\left(\tilde{u}\right)=&-\frac{1}{6}R_{0303}+\mathcal{O}\left(\gamma_{k}\right),\\
\Psi_{3}\left(\tilde{u}\right)=&-\frac{1}{2\sqrt{2}}R_{0301}+\frac{1}{2\sqrt{2}}iR_{0302}+\mathcal{O}\left(\gamma_{k}\right), \\
\Psi_{4}\left(\tilde{u}\right)=&-\frac{1}{2}R_{0101}+\frac{1}{2}R_{0202}+iR_{0102}+\mathcal{O}\left(\gamma_{k}\right), \\
\Phi_{22}\left(\tilde{u}\right)=&-\frac{1}{2}R_{0101}-\frac{1}{2}R_{0202}+\mathcal{O}\left(\gamma_{k}\right),
\end{align}
while, in term of metric perturbation, the  NP scalars take the form
\begin{align}
\Psi_{2}\left(\tilde{u}\right)=&-\frac{1}{12}\left(2h_{03,03}-h_{00,33}-h_{33,00}\right)+\mathcal{O}\left(\gamma_{k}\right),\\
\Psi_{3}\left(\tilde{u}\right)=&-\frac{1}{4\sqrt{2}}\left(h_{01,03}-h_{13,00}\right)+\frac{1}{4\sqrt{2}}i\left(h_{02,03}-h_{23,00}\right)+\mathcal{O}\left(\gamma_{k}\right),\\
\Psi_{4}\left(\tilde{u}\right)=&\frac{1}{4}\left(h_{11,00}-h_{22,00}\right)-2ih_{12,00}+\mathcal{O}\left(\gamma_{k}\right),\\
\Phi_{22}\left(\tilde{u}\right)=&\frac{1}{4}\left(h_{11,00}+h_{22,00}\right)+\mathcal{O}\left(\gamma_{k}\right)\ .
\end{align}
From Eqs.\eqref{FirstOrderTetradmassless} and \eqref{FirstOrderTetradmassive} for non-null and null geodesic congruences of GWs, we get, for a massless mode $\omega_{1}$ and a massive mode $\omega_{2}$,  at $\mathbf{k}$ fixed at the lowest order in $\gamma_{k}$, the expressions
\begin{align*}
\Psi_{2}\left(\tilde{u}\right)&=\mathcal{O}\left(\gamma_{k}\right) ,\\
\Psi_{3}\left(\tilde{u}\right)&=\mathcal{O}\left(\gamma_{k}\right)\\
\Psi_{4}\left(\tilde{u}\right)&=-\omega_{1}^{2}\left[\left(\epsilon^{(+)}\left(\omega_{1}\right)e^{i\omega_{1}\left(t-z\right)}+c.c.\right)-i\left(\epsilon^{(\times)}\left(\omega_{1}\right)e^{i\omega_{1}\left(t-z\right)}+c.c.\right)\right]+\mathcal{O}\left(\gamma_{k}\right)\\
\Phi_{22}\left(\tilde{u}\right)&=\frac{k_{z}^{2}}{8}A\left(k_{z}\right)e^{i k_{z}\left(t-z\right)}+\mathcal{O}\left(\gamma_{k}\right)+c.c.\ ,
\end{align*}
considering
\begin{equation}
\omega_{2}=k_{z}+\mathcal{O}\left(\gamma_{k}\right)\ ,
\end{equation}
in units  $c=1$. 
Hence, being $\Psi_{2}=\Psi_{3}=0$ and $\Phi_{22}\neq 0$ at zero order in $\gamma_{k}$, the quasi-Lorentz invariant $E(2)$ class of non-local gravitational theory $R+a_{1}R\Box^{-1}R$  is $N_{3}$,  according to the Petrov classification. Here  the presence or absence of all modes is observer-independent.
The driving-force matrix $S(t)$ can be expressed in terms of the six new basis polarization matrices $W_{A}(\mathbf{e}_{z})$ along the  wave direction $\hat{\mathbf{k}}=\mathbf{e}_{z}$ as 
\begin{equation}
S\left(t\right)=\sum_{A}p_{A}\left(\mathbf{e}_{z},t\right)W_{A}\left(\mathbf{e}_{z}\right)\ ,
\end{equation}
where the index $A$ ranges over $\{1,2,3,4,5,6\}$ \cite{ELL,ELLWW} and explicitly
\begin{align}
W_{1}\left(\mathbf{e}_{z}\right)=&-6\begin{pmatrix} 
0 & 0 & 0 \\
0 & 0 & 0 \\
0 & 0 & 1
\end{pmatrix}\ , & W_{2}\left(\mathbf{e}_{z}\right)=&-2\sqrt{2}\begin{pmatrix} 
0 & 0 & 1 \\
0 & 0 & 0 \\
1 & 0 & 0
\end{pmatrix}\ ,\nonumber\\
W_{3}\left(\mathbf{e}_{z}\right)=&2\sqrt{2}\begin{pmatrix} 
0 & 0 & 0 \\
0 & 0 & 1 \\
0 & 1 & 0
\end{pmatrix}\ , & W_{4}\left(\mathbf{e}_{z}\right)=&-\begin{pmatrix} 
1 & 0 & 0 \\
0 & -1 & 0 \\
0 & 0 & 0
\end{pmatrix}\ ,\nonumber\\
W_{5}\left(\mathbf{e}_{z}\right)=&\begin{pmatrix} 
0 & 1 & 0 \\
1 & 0 & 0 \\
0 & 0 & 0
\end{pmatrix}\ , & W_{6}\left(\mathbf{e}_{z}\right)=&-\begin{pmatrix} 
1 & 0 & 0 \\
0 & 1 & 0 \\
0 & 0 & 0
\end{pmatrix}\ .
\end{align}
Consequently,  there are six polarizations modes: the longitudinal mode $p_{1}^{\left(l\right)}$, the vector-$x$ mode $p_{2}^{\left(x\right)}$, the vector-$y$ mode $p_{3}^{\left(y\right)}$, the plus mode $p_{4}^{\left(+\right)}$, the cross mode $p_{5}^{\left(\times\right)}$, and the breathing mode $p_{6}^{\left(b\right)}$.
Here $p_{A}\left(\mathbf{e}_{z},t\right)$ are the amplitudes of the wave at the detector in the frame origin \cite{AMA, MY, BCLN,MN, WILL}. Taking into account that the spatial components of matrix $S(t)$ are the electric components of Riemann tensor, that is
\begin{equation}
S_{ij}(t)=R_{i0j0}\ ,
\end{equation}
we can  get the six polarization amplitudes $p_{A}\left(\mathbf{e}_{z}, t\right)$ in terms of the NP scalars  for our non-local waves
\begin{align}\label{masspolarampl}
p_{1}^{\left(l\right)}\left( \mathbf{e}_{z}, t\right)&=\Psi_{2}=\mathcal{O}\left(\gamma_{k}\right)\ ,\nonumber\\
p_{2}^{\left(x\right)}\left(\mathbf{e}_{z}, t\right)&=\text{Re}\;\Psi_{3}=p_{3}^{\left(y\right)}\left(\mathbf{e}_{z}, t\right)=\text{Im}\;\Psi_{3}=\mathcal{O}\left(\gamma_{k}\right)\ ,\nonumber\\
p_{4}^{\left(+\right)}\left( \mathbf{e}_{z}, t\right)&=\text{Re}\;\Psi_{4}=-\omega_{1}^{2}\tilde{\epsilon}^{\left(+\right)}\left(\omega_{1}\right)e^{i\omega_{1}t}+c.c.+\mathcal{O}\left(\gamma_{k}\right)\ ,\nonumber\\
p_{5}^{\left(\times\right)}\left(\mathbf{e}_{z}, t\right)&=\text{Im}\;\Psi_{4}=\omega_{1}^{2}\tilde{\epsilon}^{\left(\times\right)}\left(\omega_{1}\right)e^{i\omega_{1}t}+c.c.+\mathcal{O}\left(\gamma_{k}\right)\ ,\nonumber\\
p_{6}^{\left(b\right)}\left(\mathbf{e}_{z}, t\right)&=\Phi_{22}=\frac{k_{z}^{2}}{8}A\left(k_{z}\right)e^{i k_{z}t}+\mathcal{O}\left(\gamma_{k}\right)+c.c.\ .
\end{align}
It is clear, from Eqs.\eqref{masspolarampl},  that the two vector modes $p_{2}^{\left(x\right)}$,  and $p_{3}^{\left(y\right)}$,  together with the longitudinal scalar mode $p_{1}^{\left(l\right)}$, are suppressed at the first order in $\gamma_{k}$,  while the two standard  plus and cross  transverse tensor polarization modes, $p_{4}^{\left(+\right)}$ and $p_{5}^{\left(\times\right)}$ of frequency $\omega_{1}$, survive together with the  transverse breathing scalar mode $p_{6}^{\left(b\right)}$ of frequency  $\omega_{2}$ at zero order in $\gamma_{k}$. 

In summary, the gravitational radiation for non-local gravity $R+a_{1}R\Box^{-1}R$ shows three polarizations: $(+)$ and $(\times)$ massless 2-spin transverse tensor modes of frequency $\omega_{1}$, governed by two degrees of freedom $\epsilon^{(+)}(\omega_{1})$ and $\epsilon^{(+)}(\omega_{2})$ and a massive 0-spin transverse scalar mode of frequency $\omega_{2}$,  namely the breathing mode, governed by one d.o.f. $A(k_{z})$.  The main results of this paper are summarized in the table below
\begin{table}[h]
\begin{equation*}
\footnotesize{
\begin{array}{ccccccccc}
\toprule
\text{Constraints}  & \text{Order} & \text{Frequency} & \text{Polarization} & \text{Type} & \text{d.o.f.} & \text{Modes} & \text{Helicity} & \text{Mass}\\
&&&&& &\text{Petrov Class}&&\\
\midrule
1+a_{1}\phi_{0}-\lambda_{0}\neq6a_{1} & 2\text{th} & \omega_{1} & 2, \text{transverse} & \text{tensor} & 2 & (+),(\times) & 2 &  0 \\
\midrule
&&& &  & & N_{2} &  \\
\midrule
1+a_{1}\phi_{0}-\lambda_{0}=6a_{1} 	& 2\text{th} & \omega_{1} & 3, \text{transverse} & \text{ tensor}  & 3 & (+), (\times), (b) & 2 & 0\\
	&	&   \omega_{2} & & \text{scalar} & & N_{3} & 0 & M\\
\bottomrule
\end{array}
}
\end{equation*}
\caption{Polarizations and modes for gravitational waves  in a theory of  gravity with non-local corrections.}
\label{pol ed eli}
\end{table}

It is worth noticing that the same approach can be used also for higher-order   theories of gravity both in metric and in teleparallel formalisms.   For details see \cite{CCC1, CCC2}.

\section{Conclusions}\label{e}

The main result of this  paper is the presence of a massive scalar gravitational mode in addition to the standard massless tensor ones  in a non-local gravity theory   of the form $R+a_{1}R\Box^{-1}R$.  This model can be considered as a straightforward extension of General Relativity where a non-local correction is taken into account. In this sense, it can be considered as an Extended Theory of Gravity where the Einstein theory is a particular case \cite{Rep1,Rep2,Rep3}.

The further scalar mode is achieved  in the limit of plane waves, that is, the observer is assumed far from the wave source.  It exhibits only a transverse polarization and not a mixed one because the longitudinal polarization is suppressed if we retain only the lowest order terms in the small parameters $\gamma$ and $\gamma_{k}$.   Such parameters take into account  deviations of the waves from  exactly massless ones propagating at the light speed.  In addition,  this further massive transverse scalar mode has helicity 2 and it is  governed by one degree of freedom $A(k_{z})$.  Via the NP formalism, we have found that the gravitational waves belong to the $N_{3}$,  $E(2)$ classes of  Petrov classification.  These results have been obtained using both the  geodesic deviation  and the NP formalism.  

The approach can be easily extended to models containing  more general  terms like $R\Box^{-k}R$   which appear as effective non-local quantum corrections. Beside the renormalization and regularization of gravitational field at ultraviolet scales \cite{Modesto1,Modesto2}, these models can be  ghost-free \cite{BLM} and their infrared counterparts can be interesting at astrophysical \cite{Kostas} and cosmological scales \cite{BCFETALL} to address the dark side issues.

Finally, detecting further modes as the scalar massive one derived here is a major signature to break the degeneracy of modified theories of gravity which could be discriminated at fundamental level \cite{Lombri}. In a forthcoming paper, we will match these non-local theories with GW observations.

 \section*{Acknowledgements}
 SC is supported  by the INFN sezione di Napoli, {\it iniziative specifiche} MOONLIGHT2 and QGSKY.  
 

\end{document}